\documentclass[12pt]{iopart}

\usepackage{graphicx}
\usepackage{braket}

\expandafter\let\csname equation*\endcsname\relax
\expandafter\let\csname endequation*\endcsname\relax
\usepackage{mathtools}
\usepackage{amssymb}
\usepackage{hyperref}
\usepackage{dsfont}
\usepackage{xcolor}

\begin{document}
	
	\title[Anomalous Diffusion, Prethermalization, Particle Binding in a Flat Band System]{Anomalous Diffusion, Prethermalization, and Particle Binding in an Interacting Flat Band System}
	\author{Mirko Daumann and Thomas Dahm}
	\address{Bielefeld University, Physics Department, Postfach 100131, D-33501 Bielefeld, Germany}
	\ead{mdaumann@physik.uni-bielefeld.de}
	\vspace{10pt}
	\begin{indented}
		\item[]\today
	\end{indented}
	
	\begin{abstract}
		We study the broadening of initially localized wave packets in a quasi one-dimensional diamond ladder with interacting, spinless fermions. 
		The lattice possesses a flat band causing localization.
		We place special focus on the transition away from the flat band many-body localized case by adding very weak dispersion.
		By doing so, we allow propagation of the wave packet on significantly different timescales which causes anomalous diffusion.
		Due to the temporal separation of dynamic processes, an interaction-induced,  prethermal equilibrium becomes apparent.
		A physical picture of light and heavy modes for this prethermal behavior can be obtained within Born-Oppenheimer approximation via basis transformation of the original Hamiltonian. 
		This reveals a detachment between light, symmetric and heavy, anti-symmetric particle species.
		We show that the prethermal state is characterized by heavy particles binding together mediated by the light particles.
	\end{abstract}
	
	\section{Introduction}
	Many-body localization (MBL) has been a major topic in condensed matter and non-equilibrium physics in the last decade \cite{2015Altman,2015Nandkishore,2017Abanin,2018Alet,2019Abanin}.
	Presence of strong, quenched disorder characterizes the typical MBL phase and leads to conservation of local information and suppressed transport.
	The question under which circumstances other isolated quantum systems can form long-lasting states with preserved quantities is under debate in a wide field of physical scenarios.
	Apart from Anderson localization \cite{1958Anderson} as driving mechanism for MBL phenomena, a plethora of systems without quenched disorder have been investigated to provide some form of localization.
	After the possibility of disorder-free localization has been discussed \cite{2014Roeck1,2014Roeck2}, observations have been made in the context of glass transitions \cite{2012Carleo,2014Schiulaz,2015Horssen,2016Hickey,2017Prem}, mixtures of light and heavy particles \cite{2014Grover,2015Schiulaz,2016Yao,2019Sirker,2020Heitmann}, in systems with an extensive number of conserved quantities \cite{2017Smith1,2017Smith2}, in gauge theories \cite{2018Smith,2018Brenes,2021Karpov}, Stark potentials \cite{2019Schulz,2019Nieuwenburg,2021Scherg,2022Zisling}, in systems with strong Hilbert space fragmentation \cite{2015Papic,2020Sala,2020Khemani,2020Rakovszky} and in an extensive variety of other implementations \cite{2016He,2016Kim,2016Pino,2017Sierant,2017Mondaini,2018Lan,2018Yarloo,2019Lerose,2020Khare,2021Hart}.\\
	More recently MBL is discussed in systems which exhibit flat bands (FBs) \cite{2020Kuno,2020Roy,2021Orito,2020Danieli,2020Li,2021Vakulchyk,2022Danieli,2023Daumann,2023Swaminathan,2023Liu,2023Ahmed}.
	FBs can appear in different kind of systems and can even appear as topologically protected surface states \cite{1996Nakada,2011Heikkilae,2013Paananen,2014Kleftogiannis,2014Paananen,2015Paananen}.
	FBs may remain stable even for strong interactions \cite{2020Tilleke}.
	Like any typical Anderson-like system they show ergodicity breaking characteristics in the strongly disordered limit, but also for very weak disorder \cite{2020Kuno,2020Roy,2021Orito} (inverse Anderson transition \cite{2006Gode}) or completely without disorder \cite{2020Li,2021Vakulchyk,2022Danieli,2023Daumann}.
	A recent study provides clarification on FB MBL \cite{2022Danieli}.
	Decoupling of FB Hamiltonians reveals that conserved charges due to compact localized states in degenerate Hilbert subspaces act as effective disorder.
	Relations to mentioned systems with extensive numbers of conserved quantities and fragmented Hilbert spaces could be drawn.\\
	In this work we investigate transport in a typical system with FB.
	We show that this transport can be diffusive as well as subdiffusive depending on the flatness of the FB.
	In the limit of very weak dispersions a metastable, prethermal equilibrium emerges \cite{2018Mori}.
	Such kind of prethermal plateaus has been observed before in systems with disorder-free MBL \cite{2015Schiulaz,2015Papic,2016Yao,2017Smith2,2018Smith,2018Yarloo,2018Lan,2019Sirker}.
	We report a noteworthy correlation between FB MBL and quasi MBL in systems with light and heavy particles \cite{2016Yao} based on the prethermalization characteristics.
	We further demonstrate that the prethermal plateau is caused by the circumstance that a light particles species (occupations in dispersive band (DB) states) act as confining potential for heavy particles (FB states).
	We show that this can be understood in terms of the Born-Oppenheimer (BO) approximation \cite{1927Born}.\\
	First, we present our system with a single FB coexisting with two DBs and discuss occurring dynamics in sec. \ref{sec:model} and \ref{sec:time}.
	Anomalous diffusion and the arising prethermalization are described in \ref{sec:anom}.
	General characteristics of the prethermal state, its dependence on system parameters and a treatment in perturbation theory are covered in sec. \ref{sec:preth} and \ref{sec:pt}.
	BO approximation is carried out in sec. \ref{sec:bo} next to obtain a physical understanding of prethermalization in this model.
	The arising energy potential for heavy FB occupations within BO approximation plays an important role for the interpretation of previous results.
	This is studied in sec. \ref{sec:bo_potential}.
	\section{Anomalous Diffusion in Diamond Ladders}
	\subsection{Model}
	\label{sec:model}
\begin{figure}[h]
		\centering
		\includegraphics{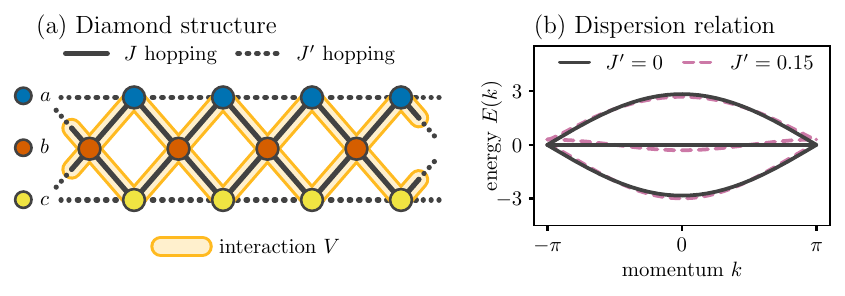}
		\caption{
			(a) Sketch and (b) single particle dispersion relation of the examined diamond ladder.}
		\label{fig:model}
	\end{figure}

	The Hamiltonian of our system (fig. \ref{fig:model}a) represents spinless fermions on a diamond lattice with periodic boundary conditions.
	This and similar structures, often also denoted as rhombic lattice, have been discussed as promising in terms of localization \cite{2000Vidal,2002Doucot,2013Hyrkas,2017Mukherjee,2018Mukherjee,2020Tilleke,2023Liu}.
	It reads:
	\begin{equation}
			\mathcal{H}=-\sum_{\langle i\sigma,j\sigma'\rangle}\left[J\cdot \hat{\mathrm{c}}^\dagger_{j\sigma'}\hat{\mathrm{c}}^{}_{i\sigma}+\frac{V}{2}\cdot\hat{\mathrm{n}}_{j\sigma'}\hat{\mathrm{n}}_{i\sigma}\right]+\sum_{\langle i\sigma,j\sigma'\rangle^*}J'\cdot\hat{\mathrm{c}}^\dagger_{j\sigma'}\hat{\mathrm{c}}^{}_{i\sigma}\ ,
			\label{eq:model}
	\end{equation}
	with fermionic creation (annihilation) operators $\hat{\mathrm{c}}^\dagger_{i\sigma}$ ($\hat{\mathrm{c}}^{}_{i\sigma}$) and occupation number operators $\hat{\mathrm{n}}_{i\sigma}=\hat{\mathrm{c}}^\dagger_{i\sigma}\hat{\mathrm{c}}^{}_{i\sigma}$.
	$\sigma\in\left\{a,b,c\right\}$ is an orbital index for the blue, red and yellow lattice sites shown in fig. \ref{fig:model}a. 
	$J$ terms on solid lines in fig. \ref{fig:model}a describe nearest neighbor hopping independent of the orbital index. 
	$J$ is also taken as unit of energy and set to $1$. 
	$J'$ on dashed lines only allows hopping between neighbored $a$ and $c$ orbitals of adjacent cells.
	\noindent
	The latter terms lead to perturbation of the FB (fig. \ref{fig:model}b) by giving it a width and thus dispersion. 
	$V$ describes a repulsive particle-particle interaction between nearest neighbors.
	$\langle i\sigma,j\sigma'\rangle$ stands for a summation over all nearest neighbored lattice sites $\{i,\sigma\}$ and $\{j,\sigma'\}$ along solid lines in fig. \ref{fig:model}a. 
		$\langle i\sigma,j\sigma'\rangle^*$ sums all connections along dashed lines between next-nearest neighbored $a$ and $c$ orbitals from adjacent cells $i$ and $j$.\\
	In the course of this article we consider three different sparsely filled systems with $L$ cells, $N=3\cdot L$ lattice sites in total, and $M \ll N$ particles. 
	The large-sized system I has $22$ cells and $M=5$ particles, the smaller system II contains $8$ cells and is filled with $M=4$ particles (except for fig. \ref{fig:sigmaV}b) and system III for interpretation purposes consists of $10$ cells and $M=3$ particles.
	The total number of particles is conserved giving the Hilbert space dimensions $d=\binom{N}{M}=(8936928,10626,4060)$ for the given systems (I, II, III).\\
	This model contains FB eigenstates inhabiting the space of anti-symmetric $a$ \& $c$ orbital superpositions due to mirror symmetry along the central axis.
	Opposite signs from anti-symmetry cause an effective decoupling of neighbored cells leading to an extensive amount of locally conserved quantities.
	This phenomenon of compact localized states is shared by several FB lattices \cite{2015Derzhko,2017Maimaiti,2018Leykam}.
	\begin{figure}[h]
		\centering
		\includegraphics{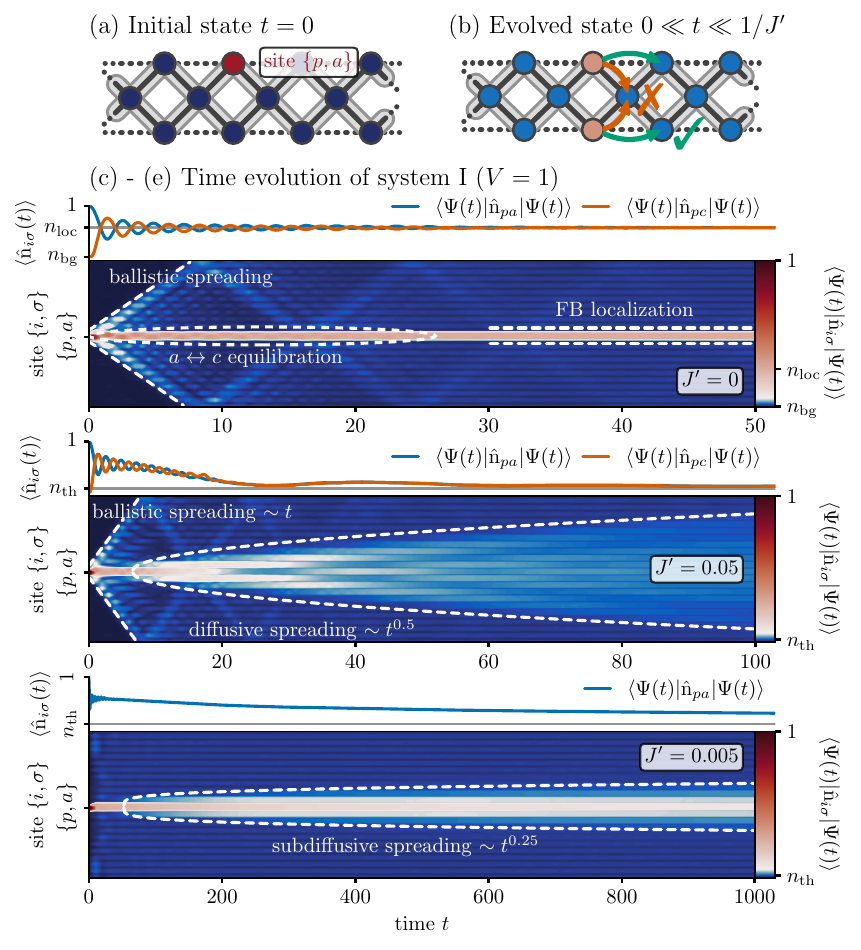}
		\caption{
			(a) Depiction of an initial wave packet with peak in particle density $\braket{\hat{\mathrm{n}}_{pa}(t=0)}=1$ at site $\{p,a\}$ (red) and random many-body background $n_{\text{bg}}\approx\frac{M-1}{N-1}$ on all other sites (dark blue). 
			(b) After inner-cell leveling between $a$ and $c$ orbital occupations to some finite value $n_{\text{loc}}$ (bright red, see top panel in (c) for comparison), particle density is constrained and only able to escape via $J'$. 
			(c) Numerical results of time evolving a single initial state in system I with $J'=0$. 
			Absence of $J'$ hopping causes FB localization recognizable by a static inhomogeneity in particle density in cell $p$. 
			(d,e) Transport away from the FB localized state can be in a diffusive or subdiffusive regime depending on the strength of $J'$. 
			Please note the different $t$ axes.}
		\label{fig:dqt}
	\end{figure}
	In addition, these eigenstates are quite robust with respect to particle-particle interaction \cite{2020Tilleke} and spread across the energy spectrum \cite{2023Daumann}.
	Local conservation caused by FB states can be bypassed by adding the perturbation $J'$, as depicted in fig. \ref{fig:dqt}b.
	
	\subsection{Time evolution}
	\label{sec:time} 
	The system is initially prepared in a highly localized state represented as a peaked wave packet.
	The propagation of this wave packet over time across the lattice is modeled with either a completely flat or slightly dispersive band. 
	$J'$ is used as parameter to continuously tune the system from a dispersive to a localized regime in order to study the transition into the FB MBL state in terms of propagation characteristics.
	The initial states fulfill the conditions of dynamical typicality \cite{2009Bartsch,2018Reimann}.
	It states that pure initial states with similar expectation values for some observable in the beginning evolve alike in that observable over time. 
	Simulating a single or only few of such states drawn at random can be considered typical and therefore representative for a much larger set of pure states.
	They are constructed in a way described in \cite{2017Steinigeweg,2017Steinigeweg2,2020Heitmann} and write:  
	\begin{equation}
		\begin{split}
			\ket{\Psi}=\hat{\mathrm{n}}_{p\tau}\cdot\ket{\Psi_r}&=\hat{\mathrm{n}}_{p\tau}\sum_{i}^{d}c_i\ket{\left\{n_{1a}\dots n_{p\tau}\dots n_{Lc}\right\}_i}\ ,	
		\end{split}
		\label{eq:is}
	\end{equation}
	where $\ket{\left\{n_{1a}\dots n_{p\tau}\dots n_{Lc}\right\}_i}$ is the full $d$-dimensional basis in occupation number representation.
	$\ket{\Psi}$'s are normalized such that $\|\Psi\|=1$.
	Projecting the particle number operator of some lattice site $\{p,\tau\}$ on a random vector $\ket{\Psi_r}$ (Gaussian distributed coefficients $c_i$ with mean $0$, variance $1$)
	generates a state where the localization maximum of the particle density forms a $\delta$ peak with $\braket{\hat{\mathrm{n}}_{i\sigma}}=1$ on site $\{p,\tau\}$.
	The other lattice sites are randomly occupied with a many-body background density $n_{\text{bg}}\approx\frac{M-1}{N-1}$.
	A depiction of such an initial state is shown as sketch on the diamond ladder in fig. \ref{fig:dqt}a. 
	Averaging of observables from several random initial states might be necessary to obtain smooth results for smaller dimensions (affecting system II).\\
	Single particle analysis shows that FB states are compositions of anti-symmetric $a$ and $c$ superpositions with empty $b$ orbitals \cite{2020Tilleke}.
	The peak of the initial wave packet is thus chosen on an $a$ orbital (interchangeable with $c$) which maximizes overlap with the FB.
	A $\delta$ occupation of an $a$ site in the single particle picture would be generated by equally adding FB and DB excitations, i.e. the FBs contribution is one half.
	This naive calculus is not accurate for many interacting particles, the idea however proves true and FB contributions are significant in $\ket{\Psi}$.\\
	First we consider time evolutions of the described states $\ket{\Psi}$ in position space.
	Calculations for system I are performed using Krylov subspace methods \cite{1950Lanczos,1986Park,2003Moler,2006Mohankumar} for matrix exponentials.
	Smaller systems II and III are fully diagonalized and the time evolution operator is created utilizing the full eigenbasis of $\mathcal{H}$.
	Fig. \ref{fig:dqt}c-e show the initial part of the equilibration process in system I for different values of $J'$ and constant interaction strength $V=1$.
	Times $t$ are given in dimensionless multiples of $J^{-1}$.
	Expectation values $\braket{\hat{\mathrm{n}}_{i\sigma}(t)}$ are calculated for a single $\ket{\Psi(t)}$.\\
	Three different dynamics can be identified: at early times $t\sim J^{-1}=1$ a ballistic, linear in $t$, spreading due to DB contributions in $\ket{\Psi}$ proceeds; during early and intermediate times oscillations between $a$ and $c$ orbitals occur until those occupations relax to some FB equilibrium value $n_{\text{loc}}$ (noticeable double peak on $\{p,a\}$ and $\{p,c\}$, see fig. \ref{fig:dqt}b and c); for $J'\neq0$, at times $t\sim J'^{-1}$ the decay of the double peak in cell $p$ sets in until some thermal value $n_{\text{th}}\approx\frac{M}{N}$ is reached on all lattice sites, see fig. \ref{fig:dqt}d and e.
	The last process is in the focus of this work.
	Its timescale and scaling with time is controlled by the $J'$ value, and thus by the width of the perturbed FB.
	Fig. \ref{fig:dqt}d shows an example for diffusive spreading caused by $J'=0.05$ while $J'=0.005$ in \ref{fig:dqt}e leads to much slower subdiffusive spreading.
	A categorization of this process into ballistic, diffusive or subdiffusive can be made based on the exponent of its power law scaling with $t$ \cite{2000Metzler,2017Lev,2019Oliveira}.
	This classification is generalized in the next step. 
	\subsection{Anomalous Diffusion}	
	\label{sec:anom}
	To get a better measure for the description of (anomalous) diffusion we introduce the time-dependent variance of the position operator (mean squared displacement) $\Sigma^2(t)$ \cite{1982Farina}. 
	It stands for the broadness of a wave packet spreading over the system. 
	The smaller $\Sigma^2(t)$ the stronger it is localized. 
	For our system it reads:
	\begin{equation}
		\begin{gathered}
			\Sigma^2(t)=\braket{\hat{\mathrm{r}}^2(t)}-\braket{\hat{\mathrm{r}}(t)}^2
			=\sum_{i=1}^L i^2\cdot n_i(t)-\left(\sum_{i=1}^L i\cdot n_i(t)\right)^2\\ 
			\text{with:}\quad n_i(t)\sim\sum_{\sigma}\left(\braket{\hat{\mathrm{n}}_{i\sigma}(t)}-n_\text{bg}\right)\ ,
		\end{gathered}
		\label{eq:sigma}
	\end{equation}
	where $n_{\text{bg}}=\frac{M-1}{N-1}$ is the many particle background density as before.
	Averaging over orbitals $\sigma$ in each cell is necessary since $a$, $b$ and $c$ for itself aren't good quantum numbers.
	The cell averaged occupations $n_i(t)$ are normalized such that $\sum_i n_i(t)=1$ at any point in time.
	Expectation values $\braket{\hat{\mathrm{n}}_{i\sigma}(t)}$ are calculated for $\ket{\Psi(t)}$ from eq. (\ref{eq:is}).
	\begin{figure}
		\centering
		\includegraphics{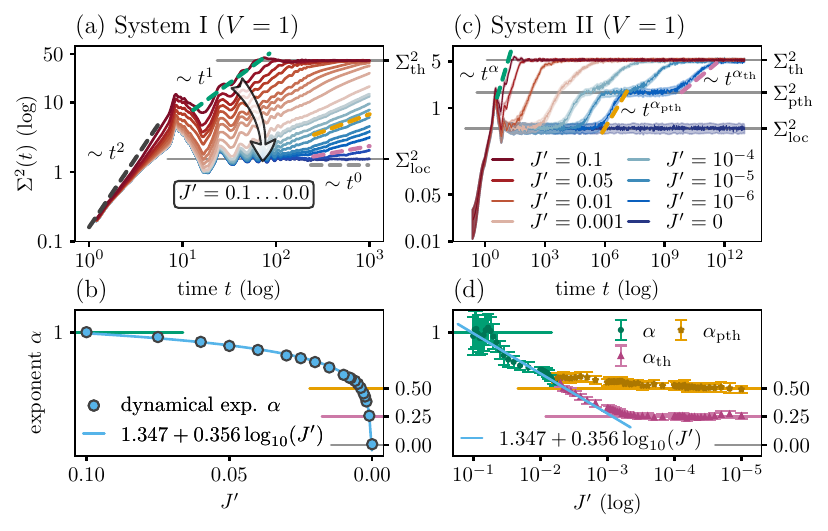}
		\caption{
			(a) Time-dependent mean squared displacement $\Sigma^2(t)$ calculated in system I for a single initial state.
			Decreasing $J'$ values slow down the propagation process. Data from fig. \ref{fig:dqt}c-e is contained in (a). 
			Some characteristic power law slopes for ballistics, diffusion, subdiffusion and complete lack of spreading are shown ($\alpha=2,1,0.5,0.25,0$).
			(b) Dynamical exponents $\alpha$ vs. $J'$ for system I. 
			Linear regression fitting of semi-log data shows a logarithmic scaling $\alpha\propto\log(J')$.
			(c) $\Sigma^2(t)$ in the $J'\rightarrow 0$ limit for long times in system II averaged for ten initial states.
			Shaded area indicate a one sigma confidence interval.
			The separation of a single $\alpha$ slope into two, $\alpha_{\text{pth}}$ and $\alpha_{\text{th}}$, with a prethermal plateau $\Sigma^2_{\text{pth}}$ in between becomes apparent for very small $J'$.
			(d) Dynamical exponents $\alpha$ vs. $J'$ for system II averaged for ten initial states with error bars. 
			$\alpha$ splits into $\alpha_{\text{pth}}$ and $\alpha_{\text{th}}$ for perturbations $J'\lesssim10^{-2}$. 
			Both settle at finite values $\lim\limits_{J'\rightarrow0}\alpha_{\text{pth}}\approx0.5$ and $\lim\limits_{J'\rightarrow0}\alpha_{\text{th}}\approx0.25$.}
		\label{fig:sigma}
	\end{figure}\\
	The course of $\Sigma^2(t)$ in fig. \ref{fig:sigma}a reproduces the dynamics shown in the time evolution of the particle density in fig. \ref{fig:dqt}c-e.
	First, we see a rapid increase from ballistic spreading due to DB contributions of the initial wave packet at short timescales followed by finite size related oscillations. 
	These stem from circulation of the density wave through the system.
	After that a dynamical regime dominated by $J'$ sets in.
	For $J'=0$, $\Sigma^2(t)$ settles down at some constant, phenomenological FB equilibrium value $\Sigma^2_{\text{loc}}$ after attenuation of the finite size oscillations.
	This state, except for small fluctuations, won't change for $t\rightarrow\infty$.
	The strength of $J'$ determines how the propagation process between $\Sigma^2_\text{loc}$ and a thermal value $\Sigma^2_\text{th}\approx\frac{1}{12}\left(L^2-1\right)>\Sigma^2_\text{loc}$ proceeds ($\Sigma^2_\text{th}$ is identical to \cite{2020Heitmann}).
	This process follows some power law $t^\alpha$ with the dynamical exponent $\alpha$ which can be extracted from slopes of the $\Sigma^2(t)$ curves in log-log graphs.
	In the present system $\alpha$ assumes values between $0\le \alpha\le 2$, which are canonically categorized as localization for $\alpha=0$, subdiffusion between $0<\alpha< 1$, normal diffusion for $\alpha=1$, superdiffusion between $1<\alpha<2$ and ballistic propagation for $\alpha=2$ \cite{2010Michalet,2013Gal,2017Steinigeweg,2017Steinigeweg2,2020Heitmann}.
	Note that $\alpha$ is double the value of given exponents in fig. \ref{fig:dqt} due to squaring of $\Sigma^2(t)$.
	Regular diffusion $\alpha\approx1$ is seen for moderately small $J'<J$. 
	For much smaller $J'\ll J$ we observe subdiffusion until a complete lack of further spreading for $J'=0$.\\
	For small $J'$ it is assumed that the simulation time frame can be cut and that observed slopes can just be extrapolated, as demonstrated in fig. \ref{fig:sigma}a. 
	The resulting exponent $\alpha$ is then shown for different $J'$ in fig. \ref{fig:sigma}b.
	Under these assumptions a smooth logarithmic scaling of $\alpha$ with $J'$ is found.
	Continuous slowdowns of propagation from normal diffusion through a subdiffusive regime to localization have been observed in studies dealing with MBL transitions in disordered systems \cite{2015Agarwal,2015Vosk,2015Potter}.
	Similar behavior has also been seen in quasiperiodic interacting systems \cite{2017Lev}.
	Logarithmic dependencies were not reported, though.
	The root of the fit function $f(x)=1.347+0.356\log_{10}(x)$ lies at $x_0=10^{-3.781}$ being the point at which a phase transition between subdiffusion and localization is to be expected.
	Yet, this result is strongly influenced by the hard cut-off at $t=10^3$ ignoring the possibility that the propagation characteristics may change at later times.
	This renders determination of the diffusion exponent $\alpha$ in the described way inaccurate.\\
	Fig. \ref{fig:sigma}a only shows a tiny segment of the whole thermalization process for small $J'$ values.
	The propagation along $J'$ doesn't set in before timescales of $\sim J'^{-1}$ are reached and the point in time $t_{\text{th}}$ (see fig. \ref{fig:sigmaV}a and c) at which $\Sigma^2(t)$ approaches $\Sigma^2_{\text{th}}$ scales with $J'^{-2}$.
	This parameter scaling with time coincides with light and heavy particle species in \cite{2016Yao}.
	In order to properly study the $J'\rightarrow0$ limit for bands with extremely weak dispersion a correspondingly long time evolution is necessary.
	This becomes impractical for systems which are not accessible via full diagonalization like system I. 
	Hence we consider the smaller system II in fig. \ref{fig:sigma}c, but for much longer times.
	At perturbation strengths $J'\lesssim10^{-3}$ visible bends in the slope from $\Sigma^2_{\text{loc}}$ to $\Sigma^2_{\text{th}}$ appear which continue to form prethermal plateaus $\Sigma^2_{\text{pth}}$ with decreasing $J'$.
	This is similar to several disorder-free systems \cite{2015Schiulaz,2015Papic,2016Yao,2017Smith2,2018Smith,2018Yarloo,2018Lan,2019Sirker} where plateaus of various observables are found in comparable cases of high parameter inequality and long time evolution.
	The duration of the system remaining in the prethermal state scales with $J'^{-2}$.
	Since the momentary state of the system at $\Sigma^2_{\text{loc}}$ and $\Sigma^2_{\text{pth}}$ violates the eigenstate thermalization hypothesis \cite{2023Daumann} and preserves information of the initial state, their values are not universal in contrast to $\Sigma^2_{\text{th}}$.\\
	$\alpha$ is only well-defined for a single slope between $\Sigma^2_{\text{loc}}$ and $\Sigma^2_{\text{th}}$ if $J'\gtrsim10^{-2}$ holds, see fig. \ref{fig:sigma}d. 
	Since emergence of $\Sigma^2_{\text{pth}}$ leads to separation of the propagation process, $\alpha$ is split into $\alpha_{\text{pth}}$ for the first slope between $\Sigma^2_{\text{loc}}$ and $\Sigma^2_{\text{pth}}$ and $\alpha_{\text{th}}$ for $\Sigma^2_{\text{pth}}$ and $\Sigma^2_{\text{th}}$. 
	When comparing both results from fig. \ref{fig:sigma}b and d with aid of fit function $f(x)=1.347+0.356\log_{10}(x)$, we find good agreement for $J'\gtrsim10^{-3}$ between $\alpha$ from system I and the combination of $\alpha$ and $\alpha_{\text{th}}$ in system II.
	Yet, in contrast to fig. \ref{fig:sigma}b where extrapolation gave a continuous decrease of $\alpha\rightarrow0$ for $J'\rightarrow0$, we observe convergence of $\alpha_{\text{pth}}$ and $\alpha_{\text{th}}$ to finite values $\lim\limits_{J'\rightarrow0}\alpha_{\text{pth}}\approx0.5$ and $\lim\limits_{J'\rightarrow0}\alpha_{\text{th}}\approx0.25$ which both lie in the subdiffusive regime.
	This model does not show a continuous FB MBL transition in terms of band flatness.
	Localization with $\alpha=0$ for $J'=0$ remains an isolated point and any perturbation $J'$, however small, causes subdiffusion and thermalization in the end.
	\section{Characteristics of the Prethermal State}
	\subsection{Dependence on System Parameters}
	\label{sec:preth}
\begin{figure}
		\centering
		\includegraphics{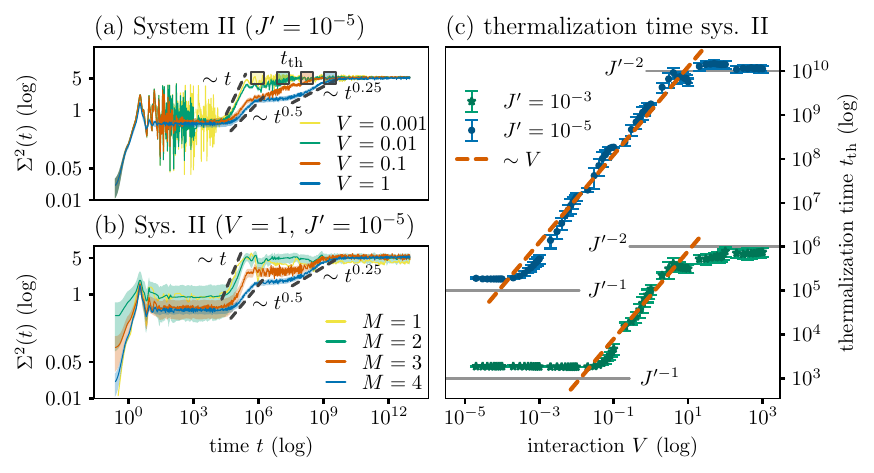}
		\caption{
			(a) $\Sigma^2(t)$ for a range of $V$ values with fixed $J'=10^{-5}$ averaged for ten initial states. 
			Shaded areas show a one sigma confidence interval. 
			Rectangles indicate determination of thermalization times $t_{\text{th}}$ for (c).
			Single particle oscillations occur at times $t\lesssim V^{-1}$. 
			Prethermalization and subdiffusion is absent for small interaction $V$.
			(b) At least $M=3$ particles are necessary to observe a prethermal plateau. 
			For $M=1$ and $M=2$ a single regular diffusive process leads to thermalization. 
			Data has been computed and averaged over $1$ ($M=1$), $1000$ ($M=2$), $100$ ($M=3$), $10$ ($M=4$) initials states. Data for $M=1$ and $M=2$ have been smoothed via Savitzky–Golay filtering \cite{1964Savitzky}.
			(c) The point in time $t_{\text{th}}$ when $\Sigma^2(t)$ approaches the thermal value $\Sigma^2_{\text{th}}$ in dependence of interaction strength $V$ for two $J'$ values.
			Data stem from averaging over ten initial states.
			Presence of interaction prolongs the thermalization process between $J'^{-1}\lesssim t_{\text{th}}\lesssim J'^{-2}$ linearly with diminishing effects in the $V\rightarrow0$ and $V\rightarrow\infty$ limits.}
		\label{fig:sigmaV}
	\end{figure}

	So far we considered a single value for interaction strength equal to the unit of energy $V=J=1$.
	Smaller values of $V$ lead to disappearance of the prethermal plateau.
	This is shown in fig. \ref{fig:sigmaV}a.
	A single diffusive process at $t\sim J'^{-1}$ instead of two subdiffusive processes leads to thermalization.
	As discussed above, $t\sim J'^{-1}$ poses a minimal time while stronger $V$ prolongs the process of thermalization up to $t\sim J'^{-2}$, see also fig. \ref{fig:sigmaV}c.
	Despite having repulsive particle-particle interaction the system tends to stay localized for longer the higher $V$ gets which has also been observed in systems with weak disorder \cite{2017Singh}.
	\\
	Variation of the total number of particles $M$ gives a very similar picture in fig. \ref{fig:sigmaV}b.
	Prethermalization and subdiffusion can only be seen for at least $M=3$ particles.
	$M=1$ and $M=2$ curves are akin to low interaction cases in fig. \ref{fig:sigmaV}a.
	This is surprising in the case of two particles in which an interacting system resembles interactionless dynamics.
	There are additional necessary conditions which occupations of flat and dispersive bands have to fulfill for prethermalization, as we will see in the following.
	\\
	To further study the influence of interaction strength $V$ on thermalization, we consider the above mentioned point in time $t_{\text{th}}$ at which the system thermalizes, as indicated in fig. \ref{fig:sigmaV}a.
	Linear scaling of $t_{\text{th}}$ with $V$ is found between $10\cdot J'\lesssim V\lesssim10\cdot J$.
	The thermalization time is prolonged between $J'^{-1}\lesssim t_{\text{th}}\lesssim J'^{-2}$.
	Together with $J'$ $t_{\text{th}}$ scales $\propto\frac{V}{J'^2}$ in this regime.
	An identical relation has been found in \cite{2016Yao} and is a noteworthy similarity between a system with FB and a mixture of light and heavy particles (spins).
	However, in the limit of weak and strong interaction its influence on the timescale ends.
	\subsection{Perturbation Theory}
	\label{sec:pt}
	Due to very different energy scales and the particularly small parameter $J'$ a treatment by means of perturbation theory is helpful. 
	A scheme for similar problems in degenerate translational-invariant spin systems is e.g. described in \cite{2018Michailidis}.
	The fact that we know exact solutions of the model for the system configuration II allows a treatment based on Brillouin-Wigner expansion series \cite{1932Brillouin} which is briefly summarized in the following.
	\begin{figure}
		\centering
		\includegraphics{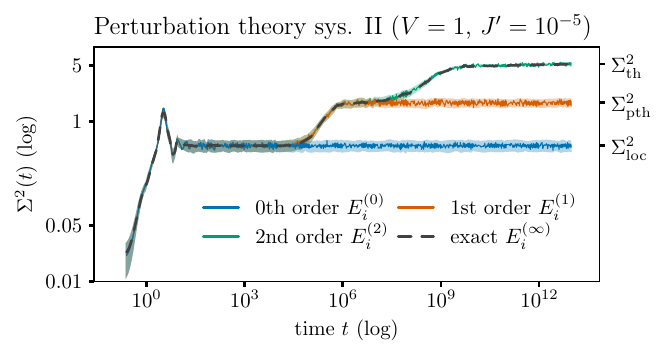}
		\caption{
			$\Sigma^2(t)$ averaged for ten initial states with shaded areas as one sigma confidence interval. 
			Without perturbation, the system remains in the FB localized state with $\Sigma^2(t)$ settling at $\Sigma^2_{\text{loc}}$. 
			The prethermal plateau $\Sigma^2_{\text{pth}}$ is reached in linear $J'$ perturbation and remains stable. 
			In second order $J'^2$ the full thermalization process can be reproduced.
		}
		\label{fig:pert}
	\end{figure}
	Perturbative $J'$ contributions are separated from the system (\ref{eq:model}) which gives two relevant Schrödinger equations:
	\begin{gather}
		\mathcal{H}_0\ket{\varphi^{(0)}_i}=E^{(0)}_i\ket{\varphi^{(0)}_i}\label{eq:pert_h0}\\
		\text{and}\nonumber\\
		\mathcal{H}\ket{\varphi_i}=\left(\mathcal{H}_0+J'\cdot\mathcal{H}_1\right)\ket{\varphi_i}=E_i\ket{\varphi_i}\label{eq:pert_h1}\ .
	\end{gather}
	$\mathcal{H}_0$ contains the diamond structure including both $J$ hopping and interactions $V$.
	$\mathcal{H}_1$ solely describes the dashed connections between neighbored cells, see fig. \ref{fig:model}a.
	$J'$ plays the role of the small parameter for perturbation series expansion.
	The time evolution operator up to the $n$'th order is constructed as follows:
	\begin{equation}
		e^{-i \mathcal{H}t}\approx\mathcal{U}\cdot\operatorname{diag}\left(e^{-i E^{(n)}_1t},\dots,e^{-i E^{(n)}_dt}\right)\cdot\mathcal{U}^{-1}\ ,
		\label{eq:time_pt}
	\end{equation}
	with $\mathcal{U}$ containing exact eigenstates $\ket{\varphi_i}$ from eq. (\ref{eq:pert_h1}).
	Thus, the order of perturbation expansion is determined by the respective order of the eigenvalues in $J'$ while eigenstates remain exact.
	The perturbed eigenvalues $E^{(n)}_i$ up to second order read:
	\begin{equation}
		E_i=E^{(\infty)}_i= \underbrace{\overbrace{E^{(0)}_i+J'\braket{\underline{\varphi}^{(0)}_i|\mathcal{H}_1|\underline{\varphi}^{(0)}_i}}^{E^{(1)}_i}+J'^2\sum_{i\neq j}\frac{\big|\braket{\underline{\varphi}^{(1)}_i|\mathcal{H}_1|\underline{\varphi}^{(1)}_j}\big|^2}{E_i-E^{(0)}_j}}_{E^{(2)}_i}+\, \mathcal{O}\big(J'^3\big)\ .
	\end{equation}
	$\ket{\underline{\varphi}^{(0)}_i}$ are identical to $\ket{\varphi^{(0)}_i}$ from eq. (\ref{eq:pert_h0}) with the only difference that degenerate subspaces $\mathcal{S}$ of $\mathcal{H}_0$ caused by the FB are diagonalized with respect to $\mathcal{H}_1$. 
	Such a procedure is common in degenerate perturbation theory (see e.g. chapter 5.2 in \cite{2020Sakurai}) and is described in the following.
	The subsequent matrix is calculated for all eigenstates $\ket{\varphi^{(0)}_{i,j}}\in\mathcal{S}$:
	\begin{equation}
			\left(\mathcal{H}^\mathcal{S}_1\right)_{i,j}=\braket{\varphi^{(0)}_{i}|\mathcal{H}_1|\varphi^{(0)}_{j}}\ .
	\end{equation}
	$\mathcal{H}^\mathcal{S}_1$ is then diagonalized by:
	\begin{equation}
		\mathcal{H}^\mathcal{S}_1\ket{\psi_i}=\varepsilon_i\ket{\psi_i}\ ,
	\end{equation}
	from which coefficients $\left(\ket{\psi_i}\right)_j$ are used to rearrange unperturbed eigenstates $\ket{\varphi^{(0)}_i}$.
	This yields the states $\ket{\underline{\varphi}^{(0)}_i}$ defined as:
	\begin{equation}
		\ket{\underline{\varphi}^{(0)}_i}=\sum_j\left(\ket{\psi_i}\right)_j\ket{\varphi^{(0)}_j}\ ,
	\end{equation}
	which are normalized due to normalization of $\ket{\psi_i}$.
	This procedure is applied on all subspaces.
	$\ket{\varphi^{(1)}_i}$ are defined as:
	\begin{equation}
		\ket{\varphi_i}=\underbrace{\ket{\underline{\varphi}^{(0)}_i}+J'\sum_{i\neq j}\frac{\braket{\underline{\varphi}^{(0)}_i|\mathcal{H}_1|\underline{\varphi}^{(0)}_j}}{E_i-E^{(0)}_j}\ket{\underline{\varphi}^{(0)}_j}}_{\ket{\varphi^{(1)}_i}}+\, \mathcal{O}\big(J'^2\big)\ ,
	\end{equation}
	of which $\ket{\underline{\varphi}^{(1)}_j}$ are again diagonalized and normalized with the formalism described above.\\
	$\Sigma^2(t)$ is shown in different orders perturbation theory in fig. \ref{fig:pert}.
	The zeroth order results are identical with the FB localized case $J'=0$.
	In first order the system evolves into the prethermal state which is stable within this approximation.
	Single hopping processes via $J'$ are incapable of leading to a further spreading of particle density.
	In second order the system evolves similar to the exact time evolution.
	It thus appears that the prethermal state is reached by first order processes in $J'$, while thermalization requires at least second order processes.
	The confining potential which traps the system in the prethermal state in first order can be found and explained in Born-Oppenheimer approximation next.
	\subsection{Born-Oppenheimer Interpretation}
	\label{sec:bo}
	The Born-Oppenheimer (BO) approximation \cite{1927Born} takes a different approach to tackle a scenario like (\ref{eq:pert_h0}) and (\ref{eq:pert_h1}) in which energy scales of system parts are magnitudes apart.
	It becomes a good approximation for high mass differences similar to perturbation theory when $J'/J$ becomes small.
	Yet, instead of solving only a single Schrödinger equation for the unperturbed part of the system and applying an expansion series to approximate the influence of the perturbation on its eigensystem, it separately solves two Schrödinger equations for both parts in parametrical dependency on each other.
	We discuss the simplest form of BO approximation for our model neglecting non-adiabatic effects.\\
	One prerequisite to utilize this approximation is the existence of differently fast modes which is clearly fulfilled in our model.
	The large energy ratio between $J$ and $J'$ results in vastly different timescales of occurring dynamics.
	The dynamics of symmetric DB modes is fast at times $\sim J^{-1}$ while anti-symmetric FB modes occur at times $\sim J'^{-1}$.
	DB states can be interpreted as a \textit{light} particle species compared to \textit{heavy} particles from FB states similar to electrons and nuclei in molecular and solid-state physics.
	Yet, while electrons and nuclei obey lepton and baryon number conservation and occupy distinct subspaces of the Hilbert space, this cannot be said for DB and FB states.
	In the present representation of (\ref{eq:model}) in fig. \ref{fig:model}a orbitals $a$ and $c$ can both be occupied by DB and FB states.
	\begin{figure}
		\centering
		\includegraphics{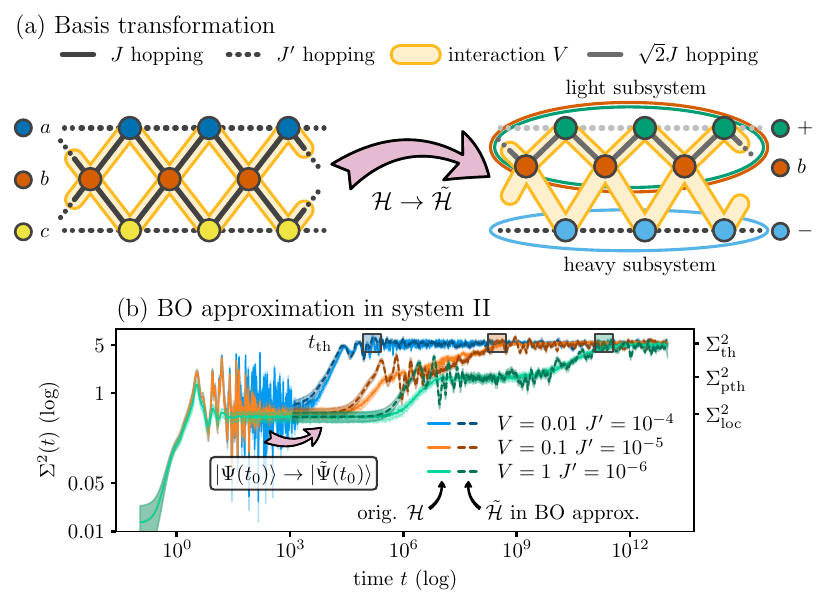}
		\caption{
			(a) Changed graph after basis transformation.
			$+$ and $b$ orbitals form the light subsystem with $\sqrt{2}J$ as main junction.
			$J'$ is neglected since $J'\ll J$.
			The heavy subsystem contains $-$ orbitals connected by $J'$.
			Both subsystems interact only via $V$ without particle interchange.
			(b) Comparison of $\Sigma^2(t)$ in the original system (solid lines) and in BO approximation after transformation (dashed lines) for a few different $V$ and $J'$ parameters. 
			$\Sigma^2(t)$ was averaged over 50 initial states with shaded area as confidence interval.
			BO data has been smoothed using \cite{1964Savitzky}.
			The courses of the original system, including prethermalization and thermalization at proper timescale, are reproduced within BO approximation with good accuracy.
			Thermalization times $t_\text{th}$ in BO approximation fit in with original data, see rectangles.
		}
		\label{fig:trafo}
	\end{figure}
	In order to formulate parametrical dependencies within BO approximation it is crucial to have two separate sets of basis states for both, light and heavy, particle species.\\
	A decoupling of light DB modes and heavy FB modes can be achieved using following canonical transformations based on their symmetry \cite{2014Flach,2022Danieli}:
	\begin{equation}
		\hat{\mathrm{d}}^{(\dagger)}_{ib}=\hat{\mathrm{c}}^{(\dagger)}_{ib}\quad\text{and}\quad\hat{\mathrm{d}}^{(\dagger)}_{i\pm}=\frac{1}{\sqrt{2}}\left(\hat{\mathrm{c}}^{(\dagger)}_{ia}\pm\hat{\mathrm{c}}^{(\dagger)}_{ic}\right)\ .
		\label{eq:trafo_operators}
	\end{equation}
	They transform the Hamiltonian $\mathcal{H}$ from eq. (\ref{eq:model}) to $\tilde{\mathcal{H}}$ with:
	\begin{equation}
		\begin{split}
			\tilde{\mathcal{H}}=-\sqrt{2}J\sum_i&\Big(\hat{\mathrm{d}}^\dagger_{i+}\hat{\mathrm{d}}^{}_{ib}+\hat{\mathrm{d}}^\dagger_{i+}\hat{\mathrm{d}}^{}_{i+1b}\Big)\\
			+\frac{V}{2}\sum_i&\Big(\hat{\mathrm{n}}_{i+}\hat{\mathrm{n}}_{ib}+\hat{\mathrm{n}}_{i+}\hat{\mathrm{n}}_{i+1b}+\hat{\mathrm{n}}_{i-}\hat{\mathrm{n}}_{ib}+\hat{\mathrm{n}}_{i-}\hat{\mathrm{n}}_{i+1b}\Big)\\
			-J'\sum_i&\Big(\hat{\mathrm{d}}^\dagger_{i+}\hat{\mathrm{d}}^{}_{i+1+}+\hat{\mathrm{d}}^\dagger_{i-}\hat{\mathrm{d}}^{}_{i+1-}\Big) + h.c. = \tilde{\mathcal{H}}_0+J'\cdot\underbrace{\left(\tilde{\mathcal{H}}_{1+}+\tilde{\mathcal{H}}_{1-}\right)}_{\tilde{\mathcal{H}}_1}\ ,
		\end{split}
		\label{eq:model_bo}
	\end{equation}
	where $\tilde{\mathcal{H}}_0$ describes the light part of the system including interaction terms and $J'\cdot\tilde{\mathcal{H}}_1$ is the heavy part.
	$\hat{\mathrm{n}}_{i\sigma}=\hat{\mathrm{d}}^\dagger_{i\sigma}\hat{\mathrm{d}}^{}_{i\sigma}$ are again occupation number operators.
	The transformation changes orbital degrees of freedom in basis states from $\sigma\in\left\{a,b,c\right\}$ to symmetry related $\sigma\in\left\{+,b,-\right\}$ and results in an altered graph shown in fig. \ref{fig:trafo}a (see also fig. 3b in \cite{2022Danieli}).
	$+$ and $b$ orbitals form a light triangular subsystem\footnote{Note that a structure like our light system itself hosts another FB for the specific values of $J'=\pm1$ which has also been discussion in terms of FB localization, see e.g. \cite{2013Hyrkas,2020Tilleke,2020Derzhko,2023Johannesmann}} in which $\sqrt{2}J$ is the dominant hopping term between sites.
	The $J'$ term in this light system is negligible due to presence of large $J$.
	Thus, $\tilde{\mathcal{H}}_{1+}$ in (\ref{eq:model_bo}) can be omitted.
	The heavy subsystem consists of $-$ orbitals with $J'$ as single connection between sites.
	Both subsystem are effectively decoupled and only interact via $V$ with each other, yet there are no transitions possible which is similar to \cite{2016Yao}.
	Occupations in the light and heavy part of the system remain conserved.
	Transforming states from the $\left\{a,b,c\right\}$ to the $\left\{+,b,-\right\}$ basis can be performed by means of the following operator:
	\begin{equation}
		\mathcal{T}=\left(\frac{1}{\sqrt{2}}\right)^\varsigma\bigotimes_{i=1}^L\hat{\mathrm{d}}^{\dagger}_{ib}\hat{\mathrm{c}}^{}_{ib}\cdot\left(\hat{\mathrm{d}}^{\dagger}_{i+}+\hat{\mathrm{d}}^{\dagger}_{i-}\right)\hat{\mathrm{c}}^{}_{ia}\cdot\left(\hat{\mathrm{d}}^{\dagger}_{i+}-\hat{\mathrm{d}}^{\dagger}_{i-}\right)\hat{\mathrm{c}}^{}_{ic}\ .
		\label{eq:transformation_op}
	\end{equation}
	$\varsigma$ is defined as the number of particles in $a$ and $c$ orbitals.\\
	The conservation of orbitals allows expansion of an eigenstate $\ket{\tilde{\varphi}_i}$ of the full transformed Hamiltonian $\tilde{\mathcal{H}}$ into the following product ansatz:
	\begin{equation}
		\ket{\tilde{\varphi}_i}=\sum_{j=1}^{d_-}\ket{\phi^+_i[\chi^-_j]}\otimes\ket{\chi^-_j}\ ,
	\end{equation}
	with $\ket{\chi^-_j}$ being a state in occupation number representation from the heavy subsystem and $\ket{\phi^+_i[\chi^-_j]}$ is a corresponding $d_+$-dimensional light eigenstate.
	The latter one parametrically depends on the configuration of heavy particles which is encoded in $\ket{\chi^-_j}$.
	Total dimensions of each subsystem for up to $M_-$ heavy and up to $M_+$ light particles are given as:
	\begin{equation}
		d_-=\sum_{m=0}^{M_-}\binom{L}{m}\quad\text{and}\quad d_+=\sum_{m=0}^{M_+}\binom{2L}{m}\ .
	\end{equation}
	The full Hilbert space dimension $d$ can be restored via $\sum\limits_{m=0}^M\binom{L}{m}\binom{2L}{M-m}=\binom{3L}{M}=d$ where $M=M_-+M_+$.\\
	The initial step to address the full $\tilde{\mathcal{H}}$ (\ref{eq:model_bo}) in BO approximation consists of first solving the light Schrödinger equation $d_-$ times:
	\begin{equation}
		\tilde{\mathcal{H}}_0\ket{\phi^+_i[\chi^-_j]}\otimes\ket{\chi^-_j}=\varepsilon^+_i[\chi^-_j]\ket{\phi^+_i[\chi^-_j]}\otimes\ket{\chi^-_j}\ ,
		\label{eq:bo_light}
	\end{equation}
	for all heavy configurations $\ket{\chi^-_j}$.
	Interaction terms in $\tilde{\mathcal{H}}_0$ cause a repulsive force between light and heavy particles which is reflected in $\varepsilon^+_i[\chi^-_j]$.
	The potential seen by the light system resembles finite $\delta$ peaks on $b$ orbitals next to heavy particle locations.
	The light system will quickly go into the groundstate assuming that the vast difference in timescales enables it to instantaneously adjust and relax with regard to any change of heavy particles configurations{\footnote{Non-adiabatic corrections would need to be considered, if heavy particle dynamics causes excitations of the light subsystem, see e.g. \cite{1955Born,2006Baer,2007Panati}. 
	Such excitations are expected to become important for increasing values of $J'/J$.}.
	Thus, for the second step of BO approximation only groundstate energies $\varepsilon^+_0[\chi^-_j]$ are taken into account.
	The second Schrödinger equation for $J'\cdot\tilde{\mathcal{H}}_1$ reads:
	\begin{equation}
		\left(\operatorname{diag}\left(\{\varepsilon^+_0[\chi^-_j]\}\right)+J'\cdot\tilde{\mathcal{H}}_1\right)\ket{\phi^-_i}=\varepsilon^-_i\ket{\phi^-_i}\ .
		\label{eq:bo_heavy}
	\end{equation}
	The light system remains untouched.
	Groundstate energies of the light system $\{\varepsilon^+_0[\chi^-_j]\}$ for all heavy configurations form a $d_-$-dimensional potential landscape through which heavy particles travel with hopping strength $J'$.\\
	A few scenarios for system II with different $V$ and $J'$ values are calculated in BO approximation in fig. \ref{fig:trafo}b by deploying the heavy eigensystem from eq. (\ref{eq:bo_heavy}) for time evolution:
	\begin{equation}
		e^{-i\tilde{\mathcal{H}}t}\approx\mathcal{U}\cdot\operatorname{diag}\left(e^{-i\varepsilon^-_1 t},\dots,e^{-i\varepsilon^-_{d_-} t}\right)\cdot\mathcal{U}^{-1}\ ,
		\label{eq:time_evolution_bo}
	\end{equation}
	where $\mathcal{U}$ contains heavy eigenstates $\ket{\phi^-_i}$.
	The initial state $\ket{\Psi(t_0)}$ for BO approximation in $\left\{a,b,c\right\}$ basis is chosen to be in the FB localized phase after relaxation of DB states at some time $t_0<J'^{-1}$.
	The operator (\ref{eq:transformation_op}) transforms $\ket{\Psi(t_0)}\rightarrow\ket{\tilde{\Psi}(t_0)}=\mathcal{T}\ket{\Psi(t_0)}$ into $\left\{+,b,-\right\}$ basis where time evolution is continued using eq. (\ref{eq:time_evolution_bo}).
	Both prethermalization and the process to a final state at thermalization time $t_\text{th}$ can be observed in this simplified model with reduced complexity and degrees of freedom.
	The dynamics in the approximated model is solely caused by the movement of heavy particles in the BO potential $\varepsilon^+_0[\chi^-_j]$ which is studied in detail next.
	\subsection{Born-Oppenheimer Potential}
	\label{sec:bo_potential}
	The potential $\varepsilon^+_0[\chi^-_j]$ originates from the interaction between the light and heavy subsystem and plays a crucial role regarding prethermalization.
	While interpretations can be extended to larger systems we restrict the following examination of $\varepsilon^+_0[\chi^-_j]$ to the simplified system III with $L=10$ cells and a total of $M=3$ particles for the sake of phenomenological discussion.
	The number of particles in each subsystem is further fixed to a single light particle, $M_+=1$, and two heavy particles, $M_-=2$.
	A single light particle is required to mediate interaction between heavy particles, while at least two of those are necessary to create nontrivial $\varepsilon^+_0[\chi^-_j]$.
	\begin{figure}
		\centering
		\includegraphics{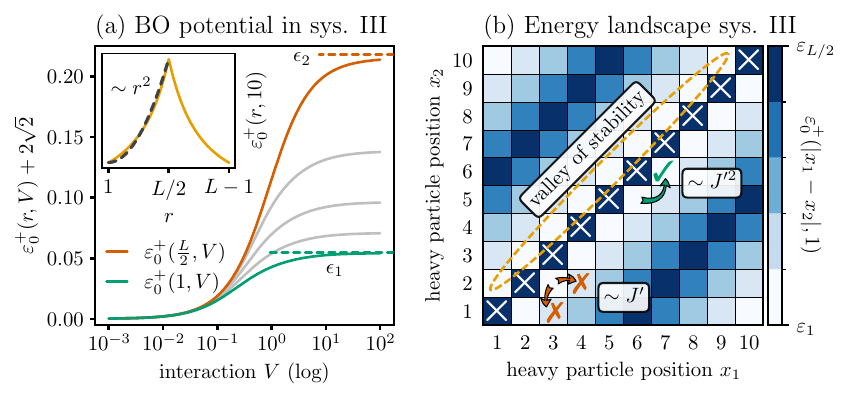}
		\caption{
			(a) Interaction-dependent values of $\varepsilon^+_0(r,V)$ for energetically most favorable $r=1$ (green) and unfavorable $r=L/2$ (red) heavy particle arrangements. 
			Configurations with intermediate values of $r$ are shown in gray.
			The non-interacting groundstate energy $2\sqrt{2}$ comes from light $\tilde{\mathcal{H}}_0$ for $V=0$.
			In the strongly interacting limit $\varepsilon^+_0[\chi^-_j]$ approaches groundstate values $\epsilon_{1}$ and $\epsilon_{2}$ of an infinite potential well.
			Inset: Dependency of $\varepsilon^+_0$ on heavy particle distance $r$ for strong $V$ which has approximately harmonic shape.
			(b) Energy landscape $\varepsilon^+_0(|x_1-x_2|,V=1)$ in dependency of heavy particle positions $x_{1,2}$ with $\varepsilon_1=\varepsilon^+_0(1,1)=-2.786$ and $\varepsilon_{L/2}=\varepsilon^+_0(L/2,1)=-2.725$. 
			All $r=|x_1-x_2|=1$ arrangements form a valley of stability which can be passed through via second order $J'^2$ hopping.}
		\label{fig:bo_energy}
	\end{figure}
	The potential for a single heavy particle would be just flat due to translational invariance of the heavy subsystem because of periodic boundary conditions.
	$\varepsilon^+_0[\chi^-_j]$ depends on interaction strength $V$ and on the relative distance $r$ between heavy particles in $\ket{\chi^-_j}$:
	\begin{equation}
		r=\frac{1}{M_-}\sum_{i\neq k}\left|\bra{\chi^-_j}i\cdot\hat{\mathrm{n}}_{i-}-k\cdot\hat{\mathrm{n}}_{k-}\ket{\chi^-_j}\right|\ ,
		\label{eq:rel_dist}
	\end{equation}
	which is why $\varepsilon^+_0(r,V)$ is used as notation from now on.\\
	Fig. \ref{fig:bo_energy}a presents a few characteristic cases for interaction and distance dependence of $\varepsilon^+_0(r,V)$.
	Considering the inset, a generic $r$-dependent course of $\varepsilon^+_0(r,V)$ for a high constant value of $V$ almost follows Hooke's law.
	It is minimal for a clustered configuration of heavy particles with $r=1$ and becomes maximized for fragmented $r=L/2$.
	In terms of interaction-dependency, BO potentials grow continuously with $V$, where $\varepsilon^+_0(1,V)<\varepsilon^+_0(2,V)<\dots<\varepsilon^+_0(L/2,V)$ holds for all $V>0$, see main part of fig. \ref{fig:bo_energy}a.
	For strong interactions limit values $\epsilon_{1}$ and $\epsilon_{2}$ are asymptotically approached.
	They can be derived from calculating the groundstate energy of the light particle being in a box since heavy particles act as infinite barriers in the $V\rightarrow\infty$ limit.
	$\epsilon_{1,2}$ read (e.g. eq. (B.12) in \cite{2020Sakurai}):
	\begin{equation}
		\epsilon_1=\frac{\sqrt{2}\pi^2}{4(L-2)^2}\quad\text{and}\quad\epsilon_2=\frac{\sqrt{2}\pi^2}{4\left(L/2-1\right)^2}\quad\text{with}\quad\epsilon_1<\epsilon_2\ .
	\end{equation}
	The mass of the particle is $1/2\sqrt{2}$ in accordance with $\tilde{\mathcal{H}}_0$, eq. (\ref{eq:model_bo}).
	The smaller $\epsilon_1$ describes a single box with interior size of $2(L-2)$, while $\epsilon_2$ is the energy of a particle distributed over two boxes with sizes $2(L/2-1)$.
	Factor $2$ takes two orbitals per unit cell into account ($+$ and $b$).
	A single wide box means lower momentum uncertainty and therefore costs less energy than multiple smaller boxes.
	The light system generally favors clustered arrangements of the heavy subsystem in terms of energy minimization.
	This is true for all $V>0$, and leads to an effective attraction between heavy particles caused by $\varepsilon^+_0(r,V)$ which we identify as mechanism behind prethermalization in this model.\\
	Another way to picture this attraction is presented in fig. \ref{fig:bo_energy}b where an energy landscape $\varepsilon^+_0(r,V)$ in dependence of both heavy particle positions $x_{1,2}$ is shown for constant $V$.
	As before there is a minimum at distance $r=|x_1-x_2|=1$ and a maximum at $r=L/2$ which are translational invariant due to periodic boundary conditions. 
	The stability of the prethermal state in first order $J'$ (fig. \ref{fig:pert}) can be explained by examining possible paths heavy particles are able to pass within this energy landscape.
	\begin{figure}
		\centering
		\includegraphics{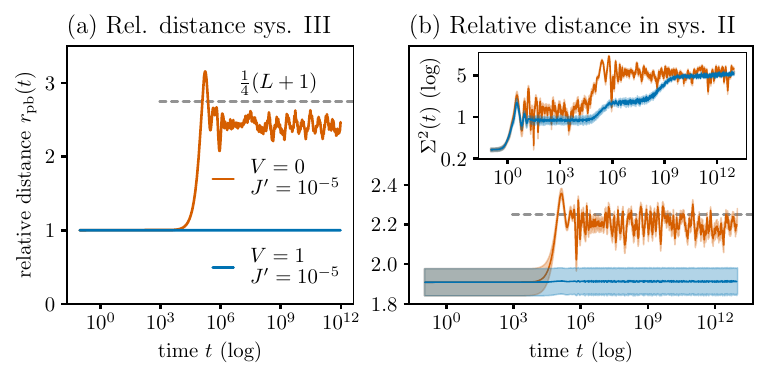}
		\caption{
			(a) Relative distance $r_\text{pb}(t)$ in BO approximation for a clustered initial state with $r=1$ in presence ($V=1$) and absence ($V=0$) of $\varepsilon^+_0(r,V)$.
			$V=0$ data has been smoothed via \cite{1964Savitzky}.
			The theoretical limit $\frac{1}{4}(L+1)$ for even distributions is marked, but not fully reached due to finite size corrections.
			(b) Relative distance $r_\text{pb}(t)$ in system II with same parameters.
			(b) shares the plot legend with (a).
			Data were averaged over 10 initial states with shaded area as confidence interval.
			$V=0$ data has been smoothed using \cite{1964Savitzky}.
			Inset: mean squared displacement $\Sigma^2(t)$ data corresponding to the main figure.}
		\label{fig:bo_rel_dist}
	\end{figure}
	In first order $J'$ two heavy particles are trapped at a single $r=1$ spot in a valley of stability if $J'$ is smaller than the potential barrier:
	\begin{equation}
		\Delta\varepsilon^+_0(V)=\varepsilon^+_0(2,V)-\varepsilon^+_0(1,V)\ .
		\label{eq:delta_eps}
	\end{equation}
	Prethermalization can thus be understood as heavy particles forming an immobile bound pair within the BO potential.
	Here we find an explanation for the necessity of at least $M=3$ particles to observe a prethermal plateau, as seen in fig. \ref{fig:sigmaV}b.
	At least two heavy particles are required for the bound pair and a single light particle is needed to mediate interactions and to create the BO potential in the first place.\\
	Second order $J'^2$ hopping, however, enables bound pairs to travel along the valley of stability what led to the observed full delocalization in terms of mean squared displacement $\Sigma^2(t)$ at times $t\sim J'^{-2}$.
	This circumstance of moving bound pairs is reflected in the time evolution of relative distance $r_\text{pb}(t)$ between heavy particles.
	It is calculated via:
	\begin{equation}
		\begin{gathered}
			r_\text{pb}(t)=\frac{1}{M_-}\sum_{i\neq k}\min(x_k(t)-x_i(t),x_i(t)-x_k(t)+L)\\ \text{where}\quad x_{i[k]}(t)=\braket{\tilde{\Psi}(t)|i[k]\cdot\hat{\mathrm{n}}_{i[k]-}|\tilde{\Psi}(t)} \quad\text{with}\quad x_i(t)<x_k(t)\ .
		\end{gathered}
		\label{eq:rel_dist_pb}
	\end{equation}
	$r_\text{pb}(t)$ is related to $r$ from eq. (\ref{eq:rel_dist}) but takes periodic boundaries into account and is defined for general states $\ket{\tilde{\Psi}(t)}$ in the transformed $\left\{+,b,-\right\}$ basis.
	Only heavy subspaces with $M_-\ge2$ are considered.
	In presence of attraction due to $\varepsilon^+_0(r,V)$, $r_\text{pb}(t)$ remains constant since movements within the valley of stability in fig. \ref{fig:bo_energy}b do not change relative distance between heavy particles.
	This is demonstrated in fig. \ref{fig:bo_rel_dist}a in an idealized case of system III in BO approximation.
	The initial state of the heavy subsystem consists of two heavy particles next to each other.
	Time evolution is performed by means of eq. (\ref{eq:time_evolution_bo}).
	$r_\text{pb}(t)$ grows in absence of $\varepsilon^+_0(r,V)$ for $V=0$ in first order $J'$ meaning that individual heavy particles are able to move freely through the system, which is in contrast to the $V=1$ case.\\
	Fig. \ref{fig:bo_rel_dist}b for system II with identical parameters confirms former results without utilizing BO approximation.
	Here, we have chosen initial states that are peaked wave packets in the original $\left\{a,b,c\right\}$ basis similar to eq. (\ref{eq:is}). 
	To account for an inhomogeneous initial distribution in terms of relative distance, they possess an additional density peak at an $a$ orbital in cell $p+1$ next to the original peak in cell $p$.
	These initial states are transformed into $\left\{+,b,-\right\}$ basis via (\ref{eq:transformation_op}) and evolved using regular time evolution of $\tilde{\mathcal{H}}$.
	Unlike before, early values of $r_\text{pb}(t)$ are larger than $1$ due to a broader distribution of the initial states caused by the random background in (\ref{eq:is}).
	Yet again, an increase of relative distance only occurs for $V=0$.
	It remains constant for $V=1$ with only minuscule fluctuations after $t>J'^{-1}$.
	Regarding the inset of fig. \ref{fig:bo_rel_dist}b, this is in contrast to $\Sigma^2(t)$ which shows the known behavior from fig. \ref{fig:sigma}a and \ref{fig:sigmaV}a.
	$\Sigma^2(t)$ increases in both cases up to thermal $\Sigma^2_\text{th}$.
	Full delocalization in terms of $\Sigma^2(t)$ occurs despite presence of the bound pairs in the heavy subsystem.
	Hence, thermalization in the parameter regime of small perturbation $J'$ and sufficiently strong $V$ does not imply breaking up of these bonds.
	This is similar to Hydrogen gas where molecules are stable at room temperature, albeit considered being thermal.\\
	This picture of heavy particle binding can also qualitatively explain the behavior of the thermalization time $t_\text{th}$ as a function of $V$ in fig. \ref{fig:sigmaV}c 
	In general, the barrier height $\Delta\varepsilon^+_0(V)$ from eq. (\ref{eq:delta_eps}) increases linearly in $V$ for $V\ll1$ and saturates to a constant value for $V\gg1$ like $\varepsilon^+_0(r,V)$ in fig. \ref{fig:bo_energy}a. 
	The crossover occurs around $V\approx1$, because $\Delta\varepsilon^+_0(V)$ is determined by the ground state of the light particles inside the static potential of a heavy particle configuration, which is proportional to $V$. 
	Concerning the thermalization time $t_\text{th}$ we can now distinguish three regimes: 1. weak interaction $V$, when $\Delta\varepsilon^+_0(V)<J'$, 2. intermediate interaction, when $\Delta\varepsilon^+_0(V)>J'$, but $V<1$, and 3. strong interaction, when $\Delta\varepsilon^+_0(V)>J'$ and $V>1$. 
	In the weak interaction regime heavy particles do not bind, because their kinetic energy $J'$ is larger than the binding energy $\Delta\varepsilon^+_0(V)$. 
	Thus, thermalization is dominated by single heavy particle hopping, which means $t_\text{th}\propto\frac{1}{J'}$. 
	In the intermediate interaction regime heavy particles bind to pairs. 
	In this case thermalization occurs via second order hopping processes through the barrier $\Delta\varepsilon^+_0(V)$, which leads to $t_\text{th}\propto\frac{\Delta\varepsilon^+_0(V)}{J'^2} \propto\frac{V}{J'^2}$.
	Finally, in the strong interaction regime $\Delta\varepsilon^+_0(V)$ becomes constant. 
	Therefore the second order hopping processes lead to $t_\text{th}\propto\frac{1}{J'^2}$.
	These three regimes and their scaling with $V$ and $J'$ are in nice agreement with the behavior seen in  fig. \ref{fig:sigmaV}c.\\
	Within the picture of heavy particle binding we can also qualitatively understand, why the dynamical exponents $\alpha_\text{pth}$ and $\alpha_\text{th}$ differ by a factor of two, as seen in fig. \ref{fig:sigma}d.
	For small values of $J'$ the thermalization of this system occurs in two well separated steps: on the timescale of $t \sim J'^{-1}$ a single heavy particle moves through the system until it meets another heavy particle to form a bound pair. 
	This process is subdiffusive with an anomalous exponent $\alpha_\text{pth}$. 
	Within the prethermal plateau the pair remains immobile until a timescale of $t \sim J'^{-2}$ is reached. 
	Thermalization finally proceeds as a correlated hopping process second order in $J'$. 
	The timescale of pair diffusion is thus squared with respect to single particle diffusion, which leads to half an exponent $\alpha_\text{th} \approx \alpha_\text{pth}/2$.
	\section{Conclusion}
	To summarize, we first studied general transport dynamics in a system with a FB which gradually receives dispersion.
	Near the FB MBL limit, i.e. FB dispersion of exactly zero, diffusion becomes slower changing into a subdiffusive regime.
	Dynamical exponents seem to continuously decrease logarithmically for weaker dispersion.
	However, when we look at fully diagonalizable systems prethermalization is revealed.
	The MBL phase remains an isolated point in this model.
	Any small dispersion causes subdiffusion and thermalization at sufficiently long times.
	The phenomenon of a prethermal state can be understood in a transformed Hamiltonian in terms of BO approximation.
	Occupations of FB states behave like heavy particles which are bound by the attractive potential mediated by light DB particles.
	This binding of heavy particles characterizes the prethermal phase.
	The BO approximation also sheds light on the thermal state which is made of delocalized bound particles in parameter regimes of strong interaction and weak dispersion.
	The behavior of the thermalization time and the ratio of two dynamical exponents can be understood as well.
	\\
	The success of BO approximation in explaining many of the observed results regarding subdiffusion and prethermalization gives opportunity for application in other disorder-free systems.
	A decoupling of degrees of freedom on an eigenstate level as done in this work, may certainly ease approaching thermodynamic limits in many models which are otherwise inaccessible.
	\section*{Acknowledgments}
	Financial support from the DFG via the research group FOR2692, grant number
	397171440 is gratefully acknowledged. We would like to thank Robin Steinigeweg, Lev Vidmar and Roderich Moessner for valuable discussions and suggestions.
	
	\section*{References}
	
	\bibliographystyle{iopart-num}
	\bibliography{anom_diff_pth_pb_fb_bib_rev}
	
\end{document}